\documentclass[%
 reprint,
 amsmath,amssymb,
 aps,
]{revtex4-2}
\usepackage{braket}
\usepackage{graphicx}
\graphicspath{ {figures/} } 
\usepackage[normalem]{ulem}

\usepackage{amsmath}
\usepackage{xcolor}
\usepackage{hyperref}
\usepackage{cleveref}
\usepackage{mathbbol}
\usepackage{bbm}

\newcommand{\ms}{\mkern-2mu}      
\newcommand{\ps}{\mkern1.7mu}    

\usepackage{xcolor}

\newcommand{\ketbra}[2]{|{#1}\rangle\!\!\,\langle{#2}|}
\newcommand{\diff}{\mathop{}\!\mathrm{d}}

\begin{document}


\title{Strong-coupling quantum optics in free space with holes in a Fermi sea}
\author{Hao Wang}
\author{Hayden C. Orth}
\author{Duo Xu}
\author{Emily J. Davis}

\affiliation{Center for Quantum Phenomena, Department of Physics, New York University, New York, NY 10003 USA}

\date{\today}

\begin{abstract}
Coherent and efficient light–matter interfaces between an atom and a single mode of the electromagnetic field are essential for quantum technologies. Traditionally, these systems employ optical cavities or waveguides to isolate a specific mode of light, but a
more recent approach is to engineer antennas from ordered configurations of trapped ultracold atoms that exhibit controllable and directed scattering. In this work, we propose a method to engineer an antenna from the center-of-mass wavefunction of a single atom, which can produce directed emission and thereby exhibit strong coupling to a target mode of light in free space. We predict that the resulting single-atom cooperativity  can be comparable to the current state of the art in optical cavity and waveguide QED experiments. Building on this approach, we show that a wavepacket antenna becomes a single-atom mirror in the linear response regime. We study the modified dipole-dipole interactions and band structure of chains of such emitters, which can exhibit sub- and superradiance at spacings much larger than the wavelength of the light. Extending our approach to multi-level atoms, we propose a method to achieve arbitrary spatial scattering from a single atom, including uni-directional spontaneous emission. Finally, we elucidate how our approach can enhance cooperative scattering in conventional arrays of tightly trapped atoms. 
\end{abstract}

\maketitle

\section{Introduction}
It is a fundamental result of classical electrodynamics that, averaging over polarization, an oscillating dipole radiates isotropically in free space. 
In the quantum optics treatment of spontaneous emission by an atom, the same result naturally emerges~\cite{cohen2024photons}. This isotropic scattering gives rise to the inherently dissipative nature of light-matter interactions: an excitation that oscillates between an atom and a target mode of the electromagnetic field must eventually leak into the continuum. Consequently, coherent and controlled light-matter interactions are notoriously challenging to engineer, yet are essential for myriad applications ranging from nonlinear optics to quantum computing and networking~\cite{kimble2008quantum, hammerer2010quantum, peyronel2012quantum, chang2014quantum, reiserer2015cavity, dhordjevic2021entanglement,sheremet2023waveguide, corzo2019waveguide}.

The conventional route to strong light-matter coupling is to define boundary conditions for the electromagnetic field, the core principle underlying both cavity and waveguide quantum electrodynamics (QED)~\cite{kimble2008quantum, hammerer2010quantum, peyronel2012quantum, chang2014quantum, reiserer2015cavity, dhordjevic2021entanglement}. Confining a target mode $M$ within a small volume can enhance the scattering rate of an emitter into that mode, $\Gamma_M$, relative to its free space scattering rate $\Gamma_{0}$. The ratio of these rates defines the cooperativity $\mathcal{C} = \Gamma_M/\Gamma_0$, the central figure of merit for light-matter interfaces. For instance, the cooperativity sets a cavity's effective photon-collection solid angle $\sim \mathcal{C}$~\cite{tanji2011interaction} and bounds the achievable circuit depth $\sim \sqrt{\mathcal{C}}$ in quantum computing applications~\cite{benito2019optimized}.

\begin{figure}[h]
    \centering
    \includegraphics[width=\linewidth]{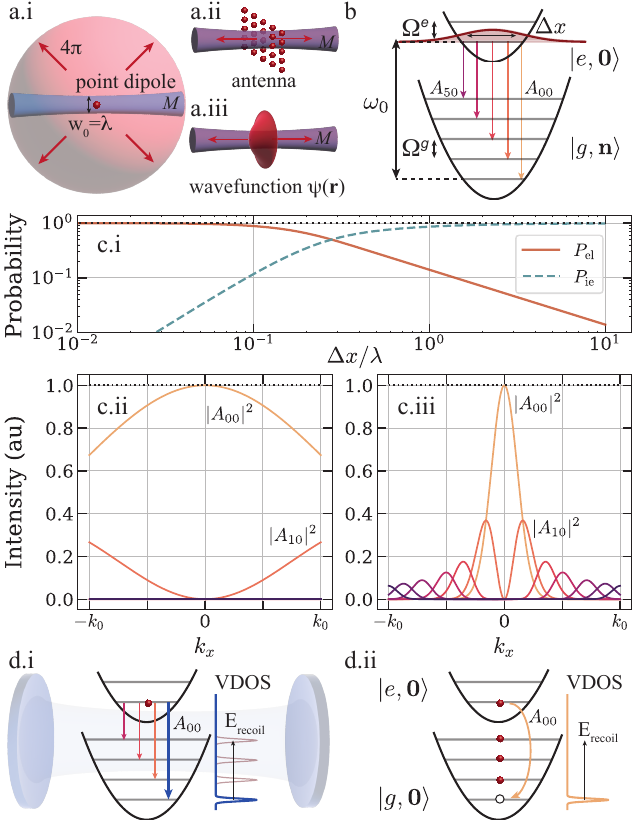}
    \caption{\textbf{Setup and overview:} (a.i) A point atom driven by mode $M$ of waist $w_0 = \lambda$ (blue) scatters photons into $4\pi$ steradians (red), with only a fraction $\mathcal{C}_\text{point} \leq 3/2\pi^2$ scattered back into mode $M$. (ii) An antenna driven by mode $M$ (blue) exhibits directed radiation (red, overlapping with blue to create violet). (iii) We study scattering from the center-of-mass wavefunction of a single atom with extent $\geq \lambda$. (b)~A single trapped atom with internal states $\{\ket{g}, \ket{e}\}$ (transition frequency $\omega_0$) and vibrational states $\mathbf{n} = (n_x, n_y, n_z)$, spaced by state-dependent trap frequencies $\mathbf{\Omega}^{g,e}$ (only one dimension is shown for clarity). An atom in the state $\ket{e, \mathbf{0}}$ scatters into $\ket{g, \mathbf{n}}$ with amplitude $A_\mathbf{n0}(\mathbf{k})$ (red arrows). (c.i) The elastic (inelastic) scattering probabilities in one dimension increase (decrease) with decreasing size of the wavefunction, $\Delta x$, compared to the wavelength $\lambda$ of the scattered photon. Inspecting the scattering amplitudes as a function of $k_x$, we observe that elastic scattering from a pointlike atom $\Delta x = \lambda/20$ is broad in $k$-space (ii), whereas a larger atom $\Delta x = \lambda$ exhibits directed elastic scattering $k_x \lesssim 1/\Delta x$ (iii). (d.i) Modifying the photon density of states with a cavity can ensure purely elastic scattering. (ii) In free space, adding identical fermions to the trap directly modifies the vibrational density of states (VDOS), and blocks all decay channels except the desired elastic channel $\ket{e,\mathbf{0}} \rightarrow \ket{g, \mathbf{0}}$. 
    }
    \label{fig:fig0}
\end{figure}

To achieve large cooperativity in free space, one can engineer an antenna. Unlike scattering from a point dipole, an ordered array of emitters can radiate into a chosen mode of the electromagnetic field and thereby reduce the dissipative coupling to the continuum. Long fundamental in fields ranging from telecommunication to nanophotonics, the principles of antenna engineering have recently been extended to ordered arrays of ultracold atoms~\cite{shahmoon2017cooperative, asenjo2017atom, rui2020subradiant, PhysRevLett.116.103602, yw5w-j13k, PhysRevLett.125.143604,barredo2018synthetic,lu2026suppression,PRXQuantum.3.010201}.  Such arrays can, in principle, achieve cooperativity scaling as  $\mathcal{C}_\text{array}\sim L^4/(\log L)^2$ when driven by a Gaussian mode whose waist is comparable to its wavelength $w_0 \approx \lambda$ and smaller than the array length $L$~[Fig.~\ref{fig:fig0}a.ii]~\cite{manzoni2018optimization}. By contrast, a single atom exhibits a relatively small cooperativity $\mathcal{C}_\text{point} \leq 3/2\pi^2$ under the same conditions~[Fig.~\ref{fig:fig0}a.i]~\cite{tanji2011interaction}. Is $\mathcal{C}_\text{point}$ a fundamental upper bound, or can an antenna comprising a single atom exhibit large cooperativity $\mathcal{C}_\text{point} \geq 1$ in free space?

In principle, one could construct an antenna from the center-of-mass wavefunction of a single atom rather than from an array of point dipoles~[Fig.~\ref{fig:fig0}a.iii]. The interference mechanism is different: an array interferes the fields of distinct emitters at fixed positions, whereas a wavepacket antenna relies on one delocalized atom interfering with itself.
Moreover, a single atom is highly nonlinear and can only absorb one photon, whereas an array is linear in the single-photon regime and indistinguishable from its classical analogue.

In practice, directed emission from a single atomic wavepacket remains unobserved due to conservation of momentum. Upon emission of a photon of momentum $\hbar k_0$, the atom experiences a recoil kick~[Fig.~\ref{fig:fig0}b]. Due to this coupling between the scattered field and the atomic center of mass, detection of the scattered photon then collapses the wavepacket. This effect becomes more pronounced for weakly trapped atoms, but such recoil kicks fundamentally limit cooperative scattering even for atoms in tight traps~[Fig.~\ref{fig:fig0}c]~\cite{rui2020subradiant, solomons2024universal, qv4c-s9xw, q2kj-w3lf, 29s3-dzl8, 82c6-4nkl, nielsen2026polaronpolaritonssubwavelengtharraystrapped,PhysRevLett.117.243601, PhysRevA.104.033718}.

In this work, we demonstrate how to circumvent the problem of photon recoil, thereby enabling directed spontaneous emission from the wavefunction of a single atom. Our approach is to block inelastic scattering channels using quantum statistics, by loading multiple fermions into a single trap~[Fig.~\ref{fig:fig0}d]. We analyze the achievable single-atom cooperativity and find that it scales nearly exponentially with fermion number~[Fig.~\ref{fig:fermions}]. We then explore the emergent long-range dipolar interactions between such emitters, which yield access to collective Dicke physics at length scales much larger than the wavelength $\lambda$~[Fig.~\ref{fig:fig4}]. Moreover, Pauli blocking can enhance coherent scattering and subradiance in arrays of tightly-trapped atoms, bypassing the fundamental limit conventionally set by the Lamb-Dicke parameter~[Fig.~\ref{fig:fig5}]~\cite{asenjo2017exponential, shahmoon2017cooperative, guimond2019subradiant, manzoni2018optimization}. Finally, we propose an extension of this method that enables nearly arbitrary spatial emission from a single atom~[Fig.~\ref{fig:fig6}]. Together, our results establish a novel pathway to access strong-coupling QED in free space.

\section{Scattering from a single trapped atom}\label{sec:sc}
To elucidate diffraction of light by a single atom, we consider spontaneous emission from an atom in a confining potential, such as an optical dipole trap. As shown in Figure~\ref{fig:fig0}b, we assume the trapped atom comprises center-of-mass $\ket{\mathbf{n}} = \ket{n_x, n_y, n_z}$ and internal $\{\ket{g}, \ket{e}\}$ degrees of freedom, where $\ket{g} \rightarrow \ket{e}$ is an electric dipole transition at frequency $\omega_0 = ck_0$. 
The bare Hamiltonian for a two-level atom trapped in a three-dimensional harmonic well is 
\begin{equation}\label{eq:H_A}
    H_A/\hbar = \omega_0\sigma_{ee}  
+ (\mathbf{\Omega}^e \cdot \hat{\mathbf{n}}^e)\sigma_{ee}  + (\mathbf{\Omega}^g \cdot \hat{\mathbf{n}}^g)\sigma_{gg},
\end{equation}
where the number operators $\hat{\mathbf{n}}^{v} = (\hat{n}_x^v, \hat{n}_y^v, \hat{n}_z^v)$ count vibrational excitations at trap frequency $\mathbf{\Omega}^{v} = (\Omega_x^v, \Omega_y^v, \Omega_z^v)$ corresponding to the internal state $v \in \{g, e\}$, and $\sigma_{vv} = \ketbra{v}{v}$. Throughout this section, we assume that the trapping potential is the same in the ground and excited states, i.e. $\mathbf{\Omega}^e = \mathbf{\Omega}^g$, although generically this may not be the case. 

Spontaneous emission from the excited state $\ket{e}$ occurs due to the electric dipole interaction between the atom and the field operators $a_\mathbf{k}$ in free space, 
\begin{equation}
H_\text{AF} = -\sum_\nu \!\int\!\mathbf{d}\cdot \boldsymbol {\mathcal{E}}_{\mathbf{k},\nu} ( \sigma_{eg} a_{\mathbf{k},\nu}  e^{i\mathbf{k}\cdot\hat{\mathbf{r}}} + \sigma_{ge} a_{\mathbf{k},\nu} ^\dagger  e^{-i\mathbf{k}\cdot\hat{\mathbf{r}}})\diff\mathbf{k},
\end{equation}
where we have dropped counter-rotating terms. 
The coupling strength is determined by the dipole operator $\hat{\mathbf{d}} = \mathbf{d}(\sigma_{ge} + \sigma_{eg})$ and the single-photon electric field $\boldsymbol {\mathcal{E}}_{\mathbf{k},\nu} $ with wavevector $\mathbf{k}$ and polarization $\nu$. Tracing out the field gives $\ket{e}$ a complex self-energy: its real part is the Lamb shift, which we absorb into $\omega_0$, and its imaginary part is the free-space decay rate $\Gamma_0 = k_0^3 |\mathbf{d}|^2/3\pi\epsilon_0\hbar$~\cite{cohen2024photons}.

The interaction $H_\text{AF}$ explicitly couples the far-field amplitude of a scattered photon to the atomic center of mass via the recoil operator $e^{-i\mathbf{k}\cdot\hat{\mathbf{r}}}$, which describes the kick imparted to the atom upon emission of a photon of momentum $\mathbf{k}$. Specifically, for an atom initially in the state $\ket{e, \mathbf{m}}$ decaying into the field vacuum $\ket{0}_\text{F}$, the amplitude to emit a photon of momentum $\mathbf{k}$ and transition to the motional state $\ket{\mathbf{n}}$ is given by $\bra{g, \mathbf{n}}\bra{1_\mathbf{k}} H_\text{AF} \ket{e, \mathbf{m}}\ket{0}_\text{F}$. This matrix element is directly proportional to the recoil coupling
\begin{equation}\label{eq:Anm}
A_{\mathbf{nm}}(\mathbf{k}) \equiv \bra{ \mathbf{n} } e^{-i\mathbf{k}\cdot \hat{\mathbf{r}}}\ket{\mathbf{m}}
\end{equation}
between motional states. The far-field amplitude of the photon emitted in this process is also proportional to $A_\mathbf{nm}$, and is thus determined by the coupling between the recoil momentum and the vibrational states of the atom.

To understand intuitively how the atomic wavefunction produces interference in the scattered field, consider an excited atom prepared in the motional ground state $\ket{e,\mathbf{m}=\mathbf{0}}$. In this case, the elastic scattering amplitude $A_\mathbf{00}(\mathbf{k})$ is the Fourier transform of  $|\psi_\mathbf{0}(\mathbf{r})|^2 = |\!\braket{\mathbf{r}|\mathbf{0}}\!|^2$, i.e.
\begin{equation}\label{eq:ft}
\langle\mathbf{0}| e^{-i\mathbf{k} \cdot \hat{\mathbf{r}}} |\mathbf{0}\rangle = \int \diff\mathbf{r}~ \psi_\mathbf{0}^*(\mathbf{r}) e^{-i\mathbf{k} \cdot \mathbf{r}} \psi_\mathbf{0}(\mathbf{r})  =\mathcal{F}[|\psi_\mathbf{0}|^2](\mathbf{k}).
\end{equation}
As the wavefunction becomes broad in real space, its Fourier transform narrows; this is precisely the antenna-like behavior of interest~[Fig.~\ref{fig:fig0}a.iii] and is reminiscent of the relationship in classical optics between the aperture function and far-field intensity~\cite{novotny2012principles}. However, the interference here arises not from distinct point sources, but from the spatial distribution of a single wavefunction. 

Unfortunately, the antenna-like behavior is completely destroyed by the recoil momentum, which causes inelastic scattering to higher vibrational states $
\ket{g, \mathbf{n} \neq \mathbf{0}}$. To illustrate this effect, we consider a one-dimensional wavefunction, $|\psi_0(x)|^2 \propto e^{-x^2/2\Delta x^2}$, whose width is comparable to the photon wavelength, $\Delta x \approx \lambda$. Because most photons are emitted with an energy $\hbar\omega_0 - E_\text{recoil} = \hbar\omega_0 - \hbar^2 k_0^2/2m$, elastic scattering is rare compared to inelastic scattering for such a large atom~[Fig.~\ref{fig:fig0}c.i]. The recoil couples strongly to higher vibrational states, up to a maximum set by the Lamb-Dicke parameter $\eta_x^2 = E_\text{recoil}/\hbar \Omega_x = (k_0 \Delta x)^2$~[Fig.~\ref{fig:fig0}c.iii]. Although elastically scattered photons are rare, their field is highly directional and concentrated in the regime $k_x \lesssim 1/\Delta x$~[Fig.~\ref{fig:fig0}c.iii, $|A_{00}|^2$ curve].

By contrast, for a pointlike atom $\eta_x \ll 1$, elastic scattering occurs with high probability~[Fig.~\ref{fig:fig0}c.i] and is nearly isotropic~[Fig.~\ref{fig:fig0}c.ii]. The rare inelastic scattering events are dominated by coupling to the first vibrational state, with probability $P_\text{ie}\approx \eta_x^2$. The tension between these two regimes -- directed but rare elastic scattering for a large atom vs. broad but frequent elastic scattering for a pointlike atom -- is fundamental. Moreover, isotropic scattering is always recovered after summing over all channels, $\sum_n |A_{n0}(k_x)|^2 = 1$, regardless of the size of the atom~[Figs.~\ref{fig:fig0}c.ii, iii, dashed black lines].

\section{Scattering into a hole in a Fermi sea}

To force a large wavepacket to exhibit only elastic spontaneous emission, it is necessary to remove the inelastic scattering channels. One approach is to modify the photonic density of states using a cavity; for example, one can overlap the cavity resonance frequency with the desired elastic decay channel as shown in Figure~\ref{fig:fig0}d.i, and suppress the emission of photons with energy less than $\hbar \omega_0$. To suppress inelastic scattering in free space, we propose engineering the atom's vibrational density of states via quantum statistics -- specifically, using Pauli blocking to eliminate unwanted dissipation. Consider a single trap filled with identical fermions. If a hole is created at the bottom of this Fermi sea by exciting a single atom $\ket{g,\mathbf{0}} \rightarrow \ket{e,\mathbf{0}}$, the excitation can only decay back to an unoccupied vibrational mode~[Fig.~\ref{fig:fig0}d.ii]. For a sufficiently large number of fermions $N$, the Fermi sea ensures that only photons with the desired far-field amplitude $A_\mathbf{00}(\mathbf{k})$ are emitted. 
Furthermore, Equation~\ref{eq:ft} shows that the radiation pattern can be engineered via the shape of the wavefunction in the trap.

This simple picture highlights how quantum statistics can induce qualitatively different scattering behavior within cold atomic ensembles, compared to individual emitters or distinguishable particles. Previous works have investigated scattering modified by quantum statistics for fermions near the Fermi surface in the trap~\cite{deb2021observation, sanner2021pauli,margalit2021pauli,o2009spontaneous}. One can also use Fermi statistics to prepare subradiant states of multilevel atoms~\cite{pineiro2020subradiance} and study the super- and subradiant physics of degenerate Fermi gases~\cite{lyne2026dicke, sandner2011spatial,bilitewski2022disentangling}.

Here, we move beyond the ensemble Fermi gas picture and explore the consequences of adding quantum statistics to light-matter interfaces that harness spatially correlated dipoles~\cite{asenjo2017exponential, PhysRevLett.117.243601, PhysRevLett.119.023603, PhysRevResearch.4.013110, PhysRevResearch.2.023086}. In particular, we show how to deterministically achieve directed spontaneous emission from a single atom. In this section, we briefly introduce the machinery for calculating scattering from multiple fermions in one trap. These tools in hand, we then analyze the directed scattering in detail.

\subsection{Scattering from fermions in a trap: Green's function approach}
We begin with a zero-temperature Fermi sea of $N$ identical fermions in a single trap, filling the $N$ lowest vibrational modes, all in the internal ground state $\ket{g}$. We denote this state as $\ket{\psi_N}$. From the Fermi sea, a single fermion in the ground motional state is then excited to prepare the target initial state $\ket{\psi_N'}$~[Fig.~\ref{fig:fig0}d.ii]. Let $S_\text{block}$ denote the set of $N-1$ vibrational modes occupied by the remaining ground-state fermions, so that we can write the two states as 
\begin{subequations}\label{eq:psi_N}
\begin{align}
    \ket{\psi_N}  &= c^\dagger_{g,\mathbf{0}}\prod_{\mathbf{n}\in S_\text{block}}
                     c^\dagger_{g,\mathbf{n}}\ket{\text{vac}}, \label{eq:psi_N_a}\\
    \ket{\psi_N'} &= c^\dagger_{e,\mathbf{0}}\prod_{\mathbf{n}\in S_\text{block}}
                     c^\dagger_{g,\mathbf{n}}\ket{\text{vac}}. \label{eq:psi_N_b}
\end{align}
\end{subequations}
The fermionic operators $c_{v,\mathbf{n}}^\dagger$ are labeled by the internal $v\in\{g, e\}$ and vibrational $\mathbf{n}$ quantum numbers; they obey the usual anti-commutation relations $\{c_{v,\mathbf{n}}^{\vphantom{\dagger}}, c_{v',\mathbf{n}\ms'}^\dagger\} = \delta_{vv'}\delta_{\mathbf{n}\mathbf{n}\ms'}$. An excited fermion cannot decay into any mode in $S_\text{block}$, which is the set of decay channels closed by Pauli blocking. The ground state $\ket{\mathbf{0}}$ remains open because it is unoccupied. 

To calculate the dynamics of the initial state $\rho = \ketbra{\psi_N'}{\psi_N'}$, we must understand the scattering from a fermionic ensemble in a single trap~\cite{lyne2026dicke, sandner2011spatial}. To this end, we turn to the dyadic Green's function operator in free space~\cite{novotny2012principles}, 
\begin{equation}\label{eq:green}
    \mathbf{G}(\mathbf{r}, \mathbf{\hat{r}'}, \omega_0) = \left(\mathbbm{1} + \frac{1}{k_0^2}\nabla \nabla \right)\frac{e^{ik_0|\mathbf{r}-\mathbf{\hat{r}'}|}}{4\pi|\mathbf{r}-\mathbf{\hat{r}'}|}.
\end{equation}
The Green's function approach provides two key advantages for the many-body problem compared to the simpler treatment of Sec.~\ref{sec:sc}. 
First, it captures the full spatial dependence of the electric field produced by a dipole. Whereas a standard far-field expansion reduces the Green's function operator to the recoil operator ($\sim e^{-i\mathbf{k} \cdot \hat{\mathbf{r}}'}$), Equation~\ref{eq:green} explicitly accounts for dipole orientation alongside near-field $1/r^3$ contributions.
Second, it enables a self-consistent calculation of the field scattered by multiple emitters.

Because the Green's function operator acts on the atomic wavefunction, the electric field at position $\mathbf{r}$ produced by a spatially-extended atom can be calculated using $\hat{\mathbf{E}}^+(\mathbf{r}, \omega_0) = \mu_0 \omega_0^2  \mathbf{G}(\mathbf{r}, \hat{\mathbf{r}}', \omega_0) \cdot \mathbf{d}\, \sigma_{ge}$,
where $\hat{\mathbf{E}}^+$ is the positive-frequency component of the field operator~\cite{lehmberg1970radiation,asenjo2017atom,dung2002resonant,gruner1996green}. 
In particular, the field produced by a transition between states $\ket{e, \mathbf{m}}$ and $\ket{g, \mathbf{n}}$ is ${\mathbf{E}}_\mathbf{nm}^+(\mathbf{r}, \omega_0)~=~\mu_0 \omega_0^2\mathbf{G}_{\mathbf{nm}}(\mathbf{r},\omega_0)~\cdot~\mathbf{d}$, where we define
\begin{align}\label{eq:Gnm}
    \mathbf{G}_{\mathbf{nm}}(\mathbf{r}, \omega_0) & \equiv \bra{\mathbf{n}} \mathbf{G}(\mathbf{r}, \hat{\mathbf{r}}', \omega_0) \ket{\mathbf{m}} \\
    & = \int \diff \mathbf{r}' \, \mathbf{G}(\mathbf{r}, \mathbf{r}', \omega_0) \psi^*_{\mathbf{n}}(\mathbf{r}') \psi_{\mathbf{m}}(\mathbf{r}').\nonumber
\end{align} 
Equation~\ref{eq:Gnm} is simply a rewriting of the Green's function operator in the basis of vibrational states instead of position, and can be understood as a weighted integral of the amplitudes $A_\mathbf{nm}(\mathbf{k})$ over all wavevectors $\mathbf{k}$ (Appendix~B).

After tracing out the field, we can write a master equation $\dot{\rho} = -i(H_\text{eff}\rho-\rho H_\text{eff}^\dagger )/\hbar + \mathcal{L}[\rho]$ for an ensemble of fermions in second-quantized form~(Appendix A)~\cite{lehmberg1970radiation,pineiro2020subradiance,asenjo2017atom,dung2002resonant,gruner1996green}. The non-Hermitian Hamiltonian $H_\text{eff}$ and Lindblad operator $\mathcal{L}$ induce dissipative spin-exchange dynamics as~\cite{sandner2011spatial}
\begin{subequations}\label{eq:second quantized H_eff}
\begin{align}
    H_{\text{eff}}^{\vphantom{\dagger}}/\hbar &= \sum_\mathbf{n}H_{A}^\mathbf{n}/\hbar - \frac{i}{2}\Gamma_0^{\vphantom{\dagger}} \sum_{\mathbf{n}} c_{e,\mathbf{n}}^\dagger c_{e,\mathbf{n}}^{\vphantom{\dagger}}\\
    &- \sum_{\mathbf{m\ms'\ms n\ms'\ms m\ps n}}\left(J_{\mathbf{m\ms'\ms n\ms'\ms m\ps n}}^{\vphantom{\dagger}} - \frac{i}{2}\gamma_{\mathbf{m\ms'\ms n\ms'\ms m\ps n}}^{\vphantom{\dagger}}\right) c_{e,\mathbf{m}\ms'}^\dagger c_{g,\mathbf{n}\ms'}^\dagger c_{g,\mathbf{m}}^{\vphantom{\dagger}} c_{e,\mathbf{n}}^{\vphantom{\dagger}}\nonumber\\
    \mathcal{L}[\rho] &= \sum_{\mathbf{m\ms'\ms n\ms'\ms m\ps n}} \gamma_{\mathbf{m\ms'\ms n\ms'\ms m\ps n}}^{\vphantom{\dagger}} c_{g,\mathbf{n}\ms'}^\dagger c_{e,\mathbf{n}}^{\vphantom{\dagger}} \rho c_{e,\mathbf{m}\ms'}^\dagger c_{g,\mathbf{m}}^{\vphantom{\dagger}}.
\end{align}
\end{subequations}
Here, we have written the dipole-dipole couplings $J, \gamma$ in the basis of vibrational modes, 
\begin{subequations}
\begin{align}
    J_{\mathbf{m\ms'\ms n\ms'\ms m\ps n}} & = \bra{\mathbf{m\ms',n\ms'}}J(\hat{\mathbf{r}}, \hat{\mathbf{r}}')\ket{\mathbf{m,n}}\\
    \gamma_{\mathbf{m\ms'\ms n\ms'\ms m\ps n}} & = \bra{\mathbf{m\ms',n\ms'}} \gamma(\hat{\mathbf{r}}, \hat{\mathbf{r}}')\ket{\mathbf{m,n}},
\end{align}
\end{subequations}
where the coherent and dissipative parts of the interaction are calculated from the Green's function as~\cite{asenjo2017atom} 
\begin{subequations}\label{eq:Jandg}
\begin{align}
    J(\hat{\mathbf{r}}, \hat{\mathbf{r}}') &= -\frac{\mu_0 \omega_0^2}{\hbar}\mathbf{d}^* \cdot \text{Re}\mathbf{G}(\hat{\mathbf{r}}, \hat{\mathbf{r}}' , \omega_0) \cdot \mathbf{d}\label{J_r}\\
    \gamma(\hat{\mathbf{r}}, \hat{\mathbf{r}}') &= \frac{2\mu_0 \omega_0^2}{\hbar}\mathbf{d}^* \cdot \text{Im}\mathbf{G}(\hat{\mathbf{r}}, \hat{\mathbf{r}}' , \omega_0) \cdot \mathbf{d}\label{Gamma_r}.
\end{align}
\end{subequations}
Equations~\ref{eq:Jandg} reduce to functions of $\mathbf{r}, \mathbf{r}'$ for the case of point dipoles~\cite{asenjo2017exponential}.

From the initial state $\rho = \ketbra{\psi_N'}{\psi_N'}$, the fermionic algebra enforces Pauli blocking, which prevents an excited fermion from decaying into an occupied level within the Fermi sea, i.e. 
$c^\dagger_{g,\mathbf{n}}c_{e,\mathbf{0}}^{\vphantom{\dagger}}\ket{\psi_N'} = c^\dagger_{g,\mathbf{n}}...c^\dagger_{g,\mathbf{n}}\ket{\text{vac}} = 0$. 
Since $\ket{\psi_N'}$ carries a single excitation $\ket{e,\mathbf{0}}$ and can only decay to an unoccupied state, the jump term $\mathcal{L}[\rho]$ reduces to
\begin{equation}
    \mathcal{L}[\rho] = \mkern -18mu \sum_{\mathbf{n},\mathbf{m} \notin S_\text{block}} \mkern -18mu\gamma_{\mathbf{0nm0}}^{\vphantom{\dagger}} \, c_{g,\mathbf{n}}^{\dagger} c_{e,\mathbf{0}}^{\vphantom{\dagger}} \rho \, c_{e,\mathbf{0}}^{\dagger} c_{g,\mathbf{m}}^{\vphantom{\dagger}}.
\end{equation}
As a result, the total decay rate from the excited state decreases with each blocked channel, producing an effective decay rate

\begin{align}\label{eq:Gamma_eff}
\Gamma_\text{eff} &= \operatorname{Tr}\mathcal{L}[\rho] = \Gamma_0 - \sum_{\mathbf{n} \in S_\text{block}} \gamma_{\mathbf{0nn0}},
\end{align}

which is reduced compared to the intrinsic linewidth $\Gamma_0$. Therefore, a reduction in the total scattering rate $\Gamma_\text{eff}~\leq~\Gamma_0$ from $\ket{\psi_N'}$ indicates Pauli blocking.

In the limit where all inelastic scattering channels are blocked $N \rightarrow \infty$, photon emission is completely restricted to the target vibrational manifold. Thus, scattering becomes purely elastic  $\Gamma_\text{el} = \gamma_{\mathbf{0000}}$, and the inelastic scattering rate $\Gamma_\text{ie} = \Gamma_\text{eff} - \Gamma_\text{el}$ goes to zero. 
In this limit, Equations~\ref{eq:second quantized H_eff} reduce to the simpler form
\begin{subequations}\label{eq:second quantized elastic H_eff}
\begin{align}
    H_{\text{eff}}^{\vphantom{\dagger}}/\hbar &= H_A^{\vphantom{\dagger}}/\hbar - \frac{i}{2}\Gamma_{\text{el}}^{\vphantom{\dagger}} c_{e,\mathbf{0}}^\dagger c_{e,\mathbf{0}}^{\vphantom{\dagger}} \\
    \mathcal{L}[\rho] &= \Gamma_{\text{el}}^{\vphantom{\dagger}} c_{g,\mathbf{0}}^\dagger c_{e,\mathbf{0}}^{\vphantom{\dagger}} \rho c_{e,\mathbf{0}}^\dagger c_{g,\mathbf{0}}^{\vphantom{\dagger}},
\end{align}
\end{subequations}
where any Stark shifts from dipole-dipole interactions have been absorbed into the bare Hamiltonian $H_A$. Here, we assume for simplicity that spin-exchange interactions can be neglected, which requires a near-magic trap satisfying $\Gamma_\text{el} \ll|\varepsilon_\Omega \Omega_s^g|  \ll \Omega_s^g$, where $\Omega_s^g$ is the radial trap frequency and we assume $\mathbf{\Omega}^e = (1+\varepsilon_\Omega)\mathbf{\Omega}^g$ (Appendix D). 
The field corresponding to the elastic scattering event is  
\begin{equation}\label{eq:E00}
     \hat{\mathbf{E}}_\mathbf{\mathbf{00}}^+(\mathbf{r},t) = \mu_0 \omega_0^2 \mathbf{G}_{\mathbf{00}}(\mathbf{r},\omega_0)\cdot \mathbf{d}\,c_{g,\mathbf{0}}^\dagger c_{e,\mathbf{0}}^{\vphantom{\dagger}}.
\end{equation}

\begin{figure}[h]
    \centering
    \includegraphics[width=\linewidth]{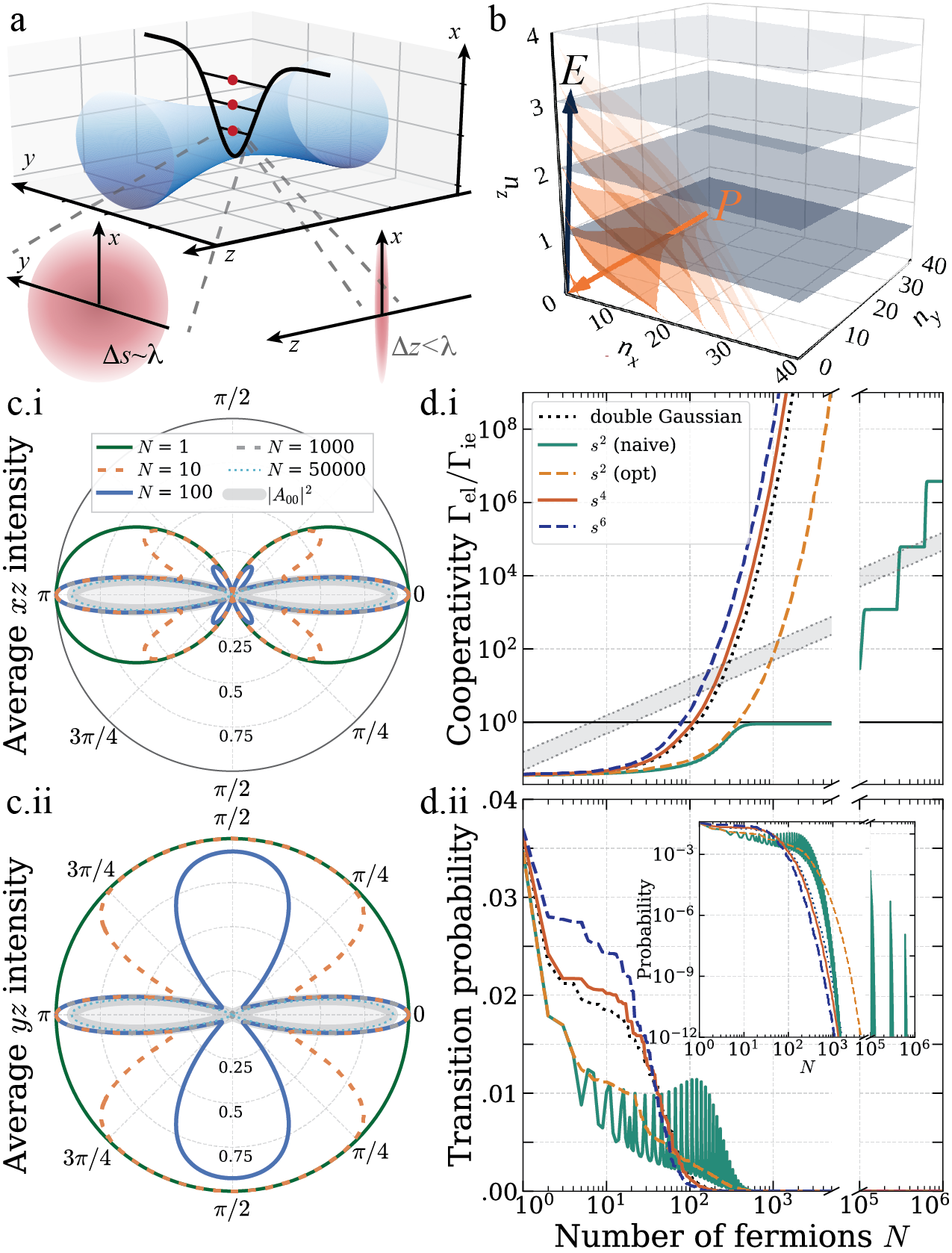}
    \vspace{-7mm}
    \caption{\textbf{Directed spontaneous emission from a disk-shaped atom:} (a) Disk-shaped wavefunction and its directed emission into a Gaussian mode propagating along the $z$-axis. (b) Iso-energy (gray) and iso-probability (orange) surfaces in phase space for $\eta_s = \sqrt{2}\pi, \eta_z =\pi/10$; surface opacity decreases with decreasing transition probability and increasing energy.  (c) The scattered intensity is plotted in the $xz$ (i) and $yz$ (ii) planes for several values of fermion number $N=1$ (solid green), $N=10$ (dotted orange), $N=100$ (solid blue), $N=1000$ (dash dot gray), and $N=50000$ (dotted cyan) along the naive loading curve in d.i. The radiation pattern approaches that of perfect elastic scattering (solid gray) as $N$ increases. (d.i) The single-atom cooperativity $\mathcal{C}_\text{disk} = \Gamma_\text{el}/\Gamma_\text{ie}$ for scattering from $\ket{e,\mathbf{0}}$ with $\eta_s = \sqrt{2}\pi, \eta_z =\pi/10$ as a function of fermions $N$ in the trap, using naive (green) vs optimized (dashed orange) loading in a harmonic potential $U \sim s^2$. Enhanced cooperativity for generalized trap shapes $U \sim s^4$ (solid orange), $s^6$ (dashed red) and a double-Gaussian (dotted blue) is also plotted. The gray shaded region shows the collective cooperativity of $N$ disordered atoms coupled to the same target mode, which ranges from $\approx .05N$ for loosely trapped or hot atoms to $.15N$ for a point dipole~(Appendix~C). Black line at $\mathcal{C}=1$ indicates the strong-coupling regime. (d.ii) The anharmonic trap shapes exhibit higher transition probabilities at lower fermion number and vice versa, compared to the harmonic trap. Inset shows log-log scale. 
    }
    \label{fig:fermions}
\end{figure}

To illustrate the effects of Pauli blocking on scattering concretely, we return briefly to the one-dimensional example discussed in Section~\ref{sec:sc}. For an initial state with two fermions, $\ket{\psi_2'} =  c^\dagger_{e,0}c^\dagger_{g,1}\ket{\text{vac}}$, radiative decay to vibrational mode $\ket{n_x = 1}$ is not permitted, and the inelastic scattering rate decreases to $\Gamma_\text{ie} = \Gamma_0 -\Gamma_\text{el} - \gamma_{0110}$. If a third fermion is added to the trap, $\ket{\psi_3'} = c^\dagger_{e,0}c^\dagger_{g,1}c^\dagger_{g,2}\ket{\text{vac}}$,  the inelastic scattering rate becomes $\Gamma_0-\Gamma_\text{el} - \gamma_{0110} - \gamma_{0220}$, and so on. Adding fermions to the ensemble can be understood as implementing vibrational selection rules, which are tuned experimentally by choosing how fermions are loaded in the trap.

\subsection{Directed spontaneous emission}\label{sec:directed}
Directed spontaneous emission from an atomic wavepacket can be achieved by combining the Pauli blocking of inelastic channels with the wavefunction design principle of Equation~\ref{eq:ft}.  Here, we study scattering into a target Gaussian mode $M$ by a wavepacket, $|\psi_\mathbf{0}(s, z)|^2 \propto e^{-s^2/2\Delta s^2}e^{-z^2/2 \Delta z^2}$, with transverse and axial Lamb-Dicke parameters $\eta_s~\gtrsim~1,~\eta_z~\ll~1$, respectively~[Fig.~\ref{fig:fermions}a]. Elastic scattering from this wavepacket produces directed emission into a solid angle $4\pi/\eta_s^2$ steradians, corresponding to a maximum polar angle of emission $\theta_\text{max} = \sqrt{2}/\eta_s$
(Appendix~B).

While the inelastic scattering rate decreases as more fermions are loaded into the trap, the elastic scattering rate is constant at fixed wavepacket size and is given by 
\begin{equation}\label{eq:Gamma_el/Gamma_0}
    \Gamma_\text{el} \approx\frac{3}{4\eta_s^2}\Gamma_0 e^{-\eta_z^2} 
\end{equation}
for $\eta_s \gtrsim 1$ and $\eta_z \ll 1$. Thus, directed emission occurs at a rate $\sim \Gamma_0/\eta_s^2$ and exhibits the same scaling with disk size as the solid angle. The decreased scattering rate $\Gamma_\text{el}\leq \Gamma_0$ should not be viewed as a drawback, because each photon is emitted precisely into the target mode $M$ without any loss to free space. In other words, emission along the $z$ axis occurs at approximately the same rate per unit solid angle as for a point emitter, and what is removed is only the unwanted emission at large angles.

By design, the cooperativity of the wavepacket antenna is simply the ratio of its elastic and inelastic scattering rates, $\mathcal{C}_\text{disk} = \Gamma_\text{el}/\Gamma_\text{ie}$; the former is spatially matched to the target mode $M$, whereas the latter is approximately orthogonal to it (Appendix C). Despite the presence of the Fermi sea, we emphasize that $\mathcal{C}_\text{disk}$ is a single-atom rather than a collective cooperativity; nonlinearity is preserved because a single excited fermion can only emit a single photon, and there is no cooperative enhancement of the scattering rate with $N$. Rather, the blocking fermions ideally do not participate in the scattering process at all.

To quantify the impact of Pauli blocking, we calculate the reduction in inelastic scattering and the corresponding increase in cooperativity as a function of the number of fermions $N$ loaded into a single disk-shaped trap.
The simplest way to fill the trap is to add fermions in order of increasing energy, corresponding to the gray iso-energy surfaces in phase space shown in Figure~\ref{fig:fermions}b. 
With wavepacket parameters $\eta_s = \sqrt{2}\pi, \eta_z =\pi/10$, and $\mathbf{d} = d\mathbf{e}_x$, the average spatial emission in the $xz$- and $yz$-planes is shown in Figure~\ref{fig:fermions}c for selected fillings $N$. As fermions are added to the trap, the scattered intensity is modified but only becomes truly directed $\theta \lesssim \sqrt{2}/\eta_s = 1/\pi$ 
for large fermion number $N \sim 10^4$. 

It follows that a large Fermi sea is required to achieve high single-atom cooperativity for this loading strategy. Indeed, the cooperativity exhibits a series of plateaus, each extending across several orders of magnitude in the fermion number $N$~[Fig.~\ref{fig:fermions}d.i, solid green curve]. As a result, the wavepacket cooperativity remains smaller than the collective cooperativity $N\mathcal{C}_\text{point}$ of $N$ disordered atoms (gray shaded reference bar) until $N \sim 10^{6}$. The plateaus arise because most photons are emitted with an energy near $\hbar\omega_0 - E_\text{recoil}$~[Fig.~\ref{fig:fig0}d]. Therefore, fermions that block improbable transitions to vibrational states well above the recoil energy play no role in suppressing emission. This inefficiency is severe in disk-shaped traps where $\eta_s \gg \eta_z$; the rapid jumps in cooperativity occur when filling small $n_x, n_y$ in the next-highest $n_z$ level. High-fidelity directed emission cannot be achieved experimentally with such poor scaling with $N$.

An efficient loading strategy is to block inelastic scattering channels in descending order of transition probability. This is equivalent to filling a pyramid in phase space up to an iso-probability surface [Fig.~\ref{fig:fermions}b, orange surfaces] set by a chosen target cooperativity~(Appendix C). A comparison of the iso-probability and iso-energy surfaces for the disk trap clearly reveals the regions of phase space where fermions are wasted. Indeed, the improved loading strategy eliminates the plateaus, and achieves approximately exponential increase of the cooperativity with $N$~[Fig.~\ref{fig:fermions}d.i, dashed orange curve]~(Appendix C). We discuss how to implement this improved loading strategy in Appendix D. 

Shaping the radial trap profile can further decrease the number of fermions required to achieve a target cooperativity. In particular, we consider super-Gaussian $-U_0e^{-(2s/w_0)^{2\alpha}}\sim s^{2\alpha}$ or double Gaussian potentials  $U_2e^{-2s^2/w_2^2}-U_1 e^{-2s^2/w_1^2} $ with $w_1 \gg w_2 \sim \Delta s$. At fixed trap depth, one can choose the anharmonic trap parameters so the directed emission lies within the same solid angle as that of the harmonic trap~(Appendix~C). All such potentials can achieve a faster increase of cooperativity with fermion number $N$ ~[Fig.~\ref{fig:fermions}d.i]. Choosing, e.g., $U \sim s^4$, the strong-coupling regime $\mathcal{C} = 1$ can be reached with $N\approx 100$ fermions; increasing to $N=300$ enables a single-atom cooperativity of 100, which is comparable with state-of-the-art optical cavity platforms~\cite{grinkemeyer2025error,PRXQuantum.4.020326, picot2026extendedsingleatomtweezerarrays}. The improved scaling with fermion number $N$ arises from the fact that the inelastic scattering amplitudes are more concentrated at lower vibrational quantum numbers for anharmonic traps~[Fig.~\ref{fig:fermions}d.ii].

Directed spontaneous emission in free space contradicts the usual intuition about scattering from a large atom when $\Gamma_0 \gg \Omega$. Typically, the recoil operator entangles the light with the atomic motion, i.e. $\sigma_{ge}a^\dagger_\mathbf{k}e^{-i\mathbf{k}\cdot \mathbf{r}}\ket{e,\mathbf{0}}\ket{0}_\text{F} = \sum_\mathbf{n} c_\mathbf{n} \ket{g,\mathbf{n}}\ket{1_\mathbf{n}}$; therefore detection of a photon with momentum $\mathbf{k}$ changes the atomic state to $\sum_\mathbf{n} c_\mathbf{n} \braket{1_\mathbf{k}|1_\mathbf{n}}\ket{g,\mathbf{n}}$. As a result, if the atom emits a photon from regions that are separated by more than the wavelength, detection of the photon yields which-path information that collapses the atomic wavefunction. Pauli blocking removes the entanglement between the atom and the photon because the atom can only decay via one channel, i.e. $c^\dagger_{g,\mathbf{0}}c_{e,\mathbf{0}}^{}a^\dagger_\mathbf{k}e^{-i\mathbf{k}\cdot \mathbf{r}}\ket{e,\mathbf{0}}\ket{0}_\text{F} = \ket{g,\mathbf{0}}\ket{1_\mathbf{0}}$, therefore detection of a photon with momentum $\mathbf{k}$ does not affect the atomic state $\braket{1_\mathbf{k}|1_\mathbf{0}}\ket{g,\mathbf{0}}$.  If $\Gamma_0 \ll \Omega$, then the state becomes more complicated (a superposition of motional coherent states), but the physics remains essentially unchanged.

\section{Applications}\label{sec:applications}
Having elucidated the unusual scattering properties of Pauli-blocked disk emitters, we present four distinct extensions of this framework. First, we show how to realize a single-atom mirror. Second, we examine dipole–dipole interactions between wavepacket antennas and analyze the resulting band structure in 1D chains. Third, we demonstrate the suppression of inelastic scattering in conventional atom arrays. Finally, we introduce a mechanism to produce an arbitrary spontaneous emission pattern from a single atom.

\subsection{Single-atom mirror}
In the linear-response regime, a wavepacket antenna behaves as a single-atom mirror. Starting with a Fermi sea $\ket{\psi_N}$ prepared in a single disk trap, we resonantly but weakly drive the $\ket{g,\mathbf{0}}\rightarrow \ket{e, \mathbf{0}}$ transition with a right-propagating paraxial Gaussian beam, 
\begin{equation}\label{eq:paraxialGaussianBeam}
\begin{split}
    \boldsymbol{\mathcal{E}}_{R}(s, z) = &  \frac{w_0}{w(z)} \mathcal{E}_0 \exp\left( \frac{-s^2}{w^2(z)} \right) \\
    & \times \exp\left( i \left[ k_0 z + \frac{k_0 s^2}{2R(z)} - \zeta(z) \right] \right) \mathbf{e}_x,
\end{split}
\end{equation}
parameterized by its Rayleigh length $z_R = \pi w_0^2/\lambda$, beam waist $w(z) = w_0 \sqrt{1 + (z/z_R)^2}$, radius of curvature $R(z)=z(1 + (z_R/z)^2)$, and Gouy phase $\zeta(z) = \arctan(z/z_R)$~\cite{siegman}. The beam waist is chosen to match the radial extent of the ground-state atomic density, $w_0 = \sqrt{2}\Delta s \geq \lambda$. Because the atomic density is symmetric about $z = 0$, the disk scatters equally to the left and right into a target mode
\begin{equation}\label{eq:EMlobes}
    \boldsymbol{\mathcal{E}}_{M}(s,z) = \begin{cases}
        \boldsymbol{\mathcal{E}}_{R}(s,z), & z\geq0,\\
        \boldsymbol{\mathcal{E}}_{L}(s,z), & z<0,
    \end{cases}
\end{equation}
where $\boldsymbol{\mathcal{E}}_{L}(s,z) = \boldsymbol{\mathcal{E}}_{R}(s,-z)$.
From Equation~\ref{eq:E00}, the field scattered by a weakly driven disk-shaped wavefunction $\eta_s\gtrsim1$, $\eta_z \ll 1$ is
\begin{equation}\label{eq:disk far field amplitude1}
   \mathbf{E}_\text{sc}(\mathbf{r}) =\frac{3\lambda^2}{8\pi^2w_0^2}\,
    \frac{i\hbar\Gamma_0}{d\,\mathcal{E}_0}\,e^{-\eta_z^2/2}
    \boldsymbol{\mathcal{E}}_M(\mathbf{r})\,
    \braket{c_{g,\mathbf{0}}^\dagger c_{e,\mathbf{0}}^{}}_\text{disk},
\end{equation}
assuming  $\braket{c_{g,\mathbf{0}}^\dagger c_{e,\mathbf{0}}^{}} \ll 1$. Since we furthermore require that $\Gamma_\text{el} \ll|\varepsilon_\Omega \Omega_s^g|$~(Appendix D), the effect of the drive $\braket{\mathcal{E}_M d}/\hbar\ll\Gamma_\text{el}$ on the remainder of the Fermi sea is far-detuned and can be neglected. 

To calculate the expectation value of the dipole operator $\braket{c_{g,\mathbf{0}}^\dagger c_{e,\mathbf{0}}^{}}_\text{disk}$, we integrate the electric field $\boldsymbol{\mathcal{E}}_M(\mathbf{r})$ over the wavefunction as
\begin{equation}\label{eq:disk polarizability}
\begin{aligned}
    \braket{c_{g,\mathbf{0}}^\dagger c_{e,\mathbf{0}}^{}}_\text{disk} 
    &= \frac{2i}{\hbar\Gamma_\text{el}}\int \diff \mathbf{r}'\,\boldsymbol{\mathcal{E}}_R(\mathbf{r}')\cdot \mathbf{d}\,|\psi_\mathbf{0}(\mathbf{r}')|^2 \\
    &= \frac{id\mathcal{E}_0}{\hbar\Gamma_\text{el}}e^{-\eta_z^2/2}.
\end{aligned}
\end{equation}
Substituting into Equation~\ref{eq:disk far field amplitude1}, the scattered field is 
\begin{equation}\label{eq:disk far field amplitude}
    \mathbf{E}_\text{sc}(s,z) =-\frac{3\lambda^2}{8\pi^2w_0^2}\frac{\Gamma_0}{\Gamma_\text{el}}e^{-\eta_z^2}\boldsymbol{\mathcal{E}}_M(s,z) = -\boldsymbol{\mathcal{E}}_M(s,z).
\end{equation}
Thus, the scattered field precisely cancels the input field at $z>0$ and reflects the light back at $z<0$; the wavepacket antenna becomes a mirror. A more general approach to derive the mirror property can be found in Appendix E.

Another way to understand the perfect reflection is by computing the scattering cross-section, $\sigma_\text{disk}$. The power scattered by the disk atom is $\hbar \omega_0 \Gamma_\text{el} \braket{c_{e,\mathbf{0}}^\dagger c_{e,\mathbf{0}}^{}}_\text{disk}$, with an excited-state population $\braket{c_{e,\mathbf{0}}^\dagger c_{e,\mathbf{0}}^{}}_\text{disk} = |\braket{c_{g,\mathbf{0}}^\dagger c_{e,\mathbf{0}}^{}}_\text{disk}|^2 = |d\mathcal{E}_0/\hbar \Gamma_\text{el}|^2 \ll 1$.
Thus the resonant cross section is 
\begin{equation}
    \sigma_\text{disk} = \sigma_\text{point}\frac{\Gamma_0}{\Gamma_\text{el}}e^{-\eta_z^2} = \frac{3\lambda^2}{2\pi}\frac{\Gamma_0}{\Gamma_\text{el}}e^{-\eta_z^2} = 4\pi w_0^2,
\end{equation}
compared to that of a point dipole $\sigma_\text{point} = 3\lambda^2/2\pi$. 
By design, the cross section of the disk emitter is simply the area of the driving mode $M$, 
which is much larger than the pointlike atom.

\subsection{Dipole-dipole interactions and band structure of disk emitters}

Because it emits light into a Gaussian spatial mode of the electromagnetic field, a wavepacket antenna behaves as though it is coupled to a waveguide despite the absence of any physical structure (e.g. an optical fiber or photonic crystal) that alters the photonic density of states.
To elucidate this behavior, we consider a one-dimensional chain of emitters whose  maximum polar angle of elastic scattering is $\theta_\text{max} =\sqrt{2}/\eta_s$. We find that the emitter spacing $a$ required to achieve strong collective effects is increased compared to the point-dipole case.
In the limit $\theta_\text{max}\rightarrow 0$, we recover precisely the physics of waveguide QED.

\begin{figure}
    \centering
    \includegraphics[width=\linewidth]{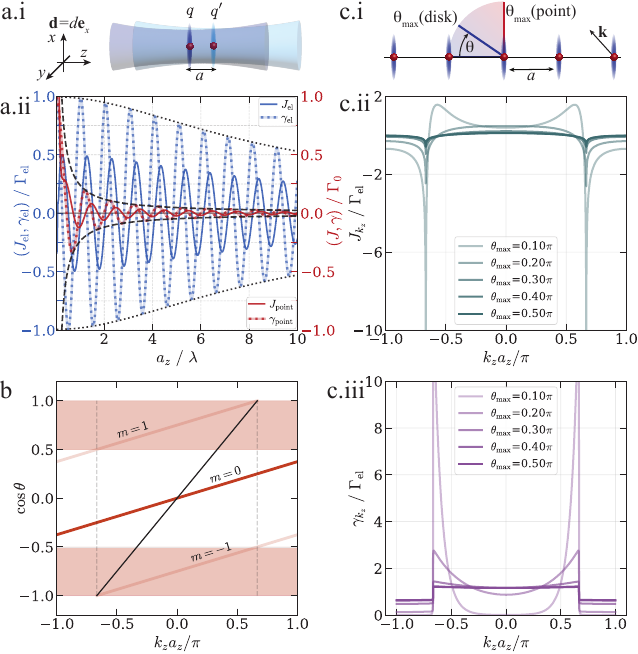}
    \caption{\textbf{Modified dipolar couplings for waveguide QED without a waveguide.} (a.i) Two disk emitters (blue) and two point emitters (red). The disks scatter into Gaussian modes whose waists are separated by the emitter spacing $a$.  (a.ii)  The pairwise dipole-dipole interactions between two emitters, showing the real part $J$ (solid) and imaginary part $\gamma$ (dotted) for disk ($\eta_s = \sqrt{2}\pi, \eta_z = \pi/10$, blue) and point (red) atoms. The disk interactions decay as a Lorentzian (dotted black), vs pointlike $1/a^3$ (dashed black). (b) Allowed diffraction orders within the light cone $|\!\cos\theta|\leq 1$ for emitter spacings $a = \lambda/3$ (black) and $a = 4\lambda/3$ (red). At subwavelength spacing $a = \lambda/3$, only one diffraction order fits inside the light cone, leaving a subradiant region $|k_z|>k_0$. For $a = 4\lambda/3$, three diffraction orders $\{-1,0,1\}$ are allowed within the conventional light cone $|\!\cos\theta|\leq 1$. However, when angles $|\theta|>\pi/3$ are excluded by directed emission, the allowed emission region shrinks (orange shaded areas) and subradiance for $|k_z|>k_0$ is recovered.  (c) Elastic band structure showing both the real $J_{k_z}$ (ii) and imaginary $\gamma_{k_z}$ (iii) Bloch mode eigenvalues for an infinite 1D chain of emitters spaced by $a = 4\lambda/3$ as a function of maximum emission angle $\theta_\text{max}$. Subradiant states both within and outside the light cone $|k_z a/\pi| = 2/3$ appear for $\theta_\text{max}/\pi = 0.1,0.2$, whereas the dissipation is nearly flat $\gamma_{k_z} \approx \Gamma_\text{el}$ for point dipoles. 
    }
    \label{fig:fig4}
\end{figure}

We first compare the dipole-dipole interactions between two disk emitters with those of two pointlike atoms~[Fig.~\ref{fig:fig4}a]. For the latter, the couplings $J, \gamma$ can be obtained directly from the Green's function~[Eqs.~\ref{eq:green},~\ref{eq:Jandg}] and exhibit the expected scaling $1/a^3$ in the near-field~[Fig.~\ref{fig:fig4}a.ii, red curves]. To obtain the dipole-dipole coupling for two disk emitters, we evaluate the Green's function restricted to the ground vibrational states as 
\begin{subequations}\label{eq:disk-disk-interaction}
\begin{align}
    J_\text{el} - \frac{i}{2}\gamma_\text{el} &= \int\!\!\!\int\!\diff\mathbf{r} \diff\mathbf{r'} \! \left[J(\mathbf{r}, \mathbf{r}') - \frac{i}{2}\gamma (\mathbf{r}, \mathbf{r}')\right]\\\nonumber
    &\hspace{10mm}\times |\!\braket{\mathbf{0}|\mathbf{r'}-a\mathbf{e_z}}\!|^2|\!\braket{\mathbf{r}|\mathbf{0}}\!|^2\\
    &\hspace{-15mm}\approx -\frac{3i\lambda}{8\pi} \frac{\Gamma_0}{\sqrt{a^2 + 4z_R^2}}  \text{exp}\left[-\eta_z^2 + i (k_0 a - \xi(a)) \right], 
\end{align}
\end{subequations}
where $z_R = 2\pi\Delta s^2/\lambda, \xi(a) = \arctan(a/2z_R)$ and we have assumed $\eta_s \gtrsim 1, \eta_z \ll 1$.
Equation~\ref{eq:disk-disk-interaction} shows that the coupling between the two disks is well-approximated by the overlap between the two Gaussian modes corresponding to their far-field emission patterns, which is itself a Gaussian mode~(Appendix E). The overall strength of the interaction is reduced by the ratio of elastic scattering $\Gamma_\text{el}$ compared to $\Gamma_0$, and the spatial dependence is converted from the conventional $1/r^3$ scaling of point dipoles to a Lorentzian profile~[Fig.~\ref{fig:fig4}a.ii, blue curves]. 
For highly directed emission ($\eta_s \gg 1$), the interactions weaken and lose their dependence on the spacing $a$.

We now form a one-dimensional chain of emitters~[Fig.~\ref{fig:fig4}c.i], for which one can derive an effective spin-exchange Hamiltonian (Appendix A),
\begin{align}\label{eq:Hspin}
    H_\text{spin}/\hbar &= \sum_{q\neq q'}\left( J_\text{el}^{qq'}  -\frac{i}{2}\gamma_\text{el}^{qq'}\right)\sigma_q^\dagger \sigma_{q'}^{} - \frac{i}{2} \Gamma_\text{eff}\sum_q \sigma_q^\dagger \sigma_{q}^{},
\end{align}
in the single-excitation manifold. Each site $q$ comprises a fermionic ensemble with dynamics in only the lowest vibrational state, i.e. the operator $\sigma_q^{\dagger} = c_{e,\mathbf{0},q}^\dagger c_{g,\mathbf{0},q}$ excites the lowest-energy fermion in disk $q$ from $\ket{g,\mathbf{0}}$ to $\ket{e,\mathbf{0}}$. To restrict the dynamics to the ground vibrational manifold, we further assume inelastic scattering is negligible and that the necessary conditions are met to suppress any dynamics within or between Fermi seas, such that $\Gamma_\text{eff} = \Gamma_\text{el}$ (Appendix D). 

To understand collective scattering from this system, we study the Bloch modes $|k_z| \leq \pi/a$ of the chain as a function of directedness of the emission. Specifically, 
we allow the size of the transverse wavefunction, and hence the maximum polar angle of emission $\theta_\text{max}$, to vary uniformly for each emitter in the chain~[Fig.~\ref{fig:fig4}c.i]. 
Intuitively, the total scattered field from Bloch mode $k_z$ is 
$\propto \sum_q \phi_q = \sum_q \exp[i(q k_0 a\cos\theta-q k_z a)]$ along the wavevector $\mathbf{k} = k_0 \sin\theta \mathbf{e}_s + k_0 \cos\theta \mathbf{e}_z$. 
Both sub- and superradiant modes $k_z$ can arise, exhibiting either suppressed or enhanced decay via destructive or constructive interference, respectively, between emitters. The dispersive shifts and linewidths $J_{k_z}, \gamma_{k_z}$, respectively, of the Bloch modes are plotted in Figure~\ref{fig:fig4}c.ii, iii for several choices of $\theta_\text{max}$.

For a subradiant mode $\gamma_{k_z} \approx 0$ of the chain to exist, the phases $\phi_q$ must destructively interfere for all emission angles $\theta < \theta_\text{max}$. For pointlike atoms, a subradiant mode exists if and only if the phases destructively interfere, $k_0a\cos\theta -~k_za~\neq~2m\pi,~m\in\mathbb{Z}$ for all $\theta\leq\pi/2$.
This condition can be satisfied only when $a <\lambda/2$, i.e. when there is only a single diffraction order $m=0$ inside the light cone $|\!\cos\theta|\leq1$ of free space modes~[Fig.~\ref{fig:fig4}b]~\cite{asenjo2017exponential}. Due to the diffraction limit of light, subwavelength spacing is extremely challenging to achieve experimentally for atoms that are optically trapped. 

By contrast, directed emission requires destructive interference only within angles $\theta \in [0, 1/\eta_s]$ to achieve subradiance, thereby enabling larger spacings $a$.
The two cases are compared in Figure~\ref{fig:fig4}b.
At spacing $a = \lambda/3$, only a single diffraction order $m=0$  (black line) exists within the light cone $|\!\cos\theta|\leq1$ and modes $|k_z| > k_0$ are subradiant (vertical gray dashed lines). For a larger spacing $a = 4\lambda/3$, there are three diffraction orders $m\in\{-1,0,1\}$ within the light cone (red lines). However, choosing a more restrictive $\theta_\text{max} = \pi/3$ hollows out the center of the light cone such that emission to the $m=0$ diffraction order is forbidden, while $m=\pm1$ are still allowed~[Fig.~\ref{fig:fig4}b, orange shaded regions]. Thus, as $\theta_\text{max}$ decreases, the hollow light cone enables subradiance to be achieved even when the chain emits into higher diffraction orders. In the limit $\theta_\text{max} \rightarrow 0$, a subradiant mode can exist at any spacing $a$ because the condition for destructive interference is simply $k_0 a - k_z a \neq 2m\pi$.

The same behavior is exhibited in Figure~\ref{fig:fig4}c, which shows the dispersion and linewidths of the Bloch modes for a chain with spacing $a = 4\lambda/3$ as a function of the emission angle $\theta_\text{max}$. Pointlike atoms are nearly independent at this spacing, resulting in a flat dispersion $J_{k_z}\approx 0$ and scattering rate $\gamma_{k_z}\approx\Gamma_\text{el}$. As the emission becomes more directed, the dispersion becomes sharply peaked at the light cone boundary $|k_z| = k_0$ (dark to light teal curves). The sharply peaked dispersion indicates the emergence of two superradiant modes $\gamma_{k_0}\sim L\Gamma_\text{el}$ (lightest violet curve). Concomitantly, subradiant modes emerge both inside and outside the light cone as the angle $\theta_\text{max}$ decreases (dark to light violet curves). These are precisely the states one would expect 
for emitters coupled to a one-dimensional waveguide~\cite{dicke1954coherence, albrecht2019subradiant}.

Collecting these examples into a coherent framework, the maximum spacing $a$ at which strong collective effects including subradiance can occur is $a\approx 2\pi z_R$ for sufficiently directed scattering such that the emission mode $M$ of each disk is paraxial. The maximum spacing is set by the Rayleigh length of the emission mode rather than its wavelength. Thus, designing directed emitters expands the phenomenon of subradiance to a much wider range of spacings and obviates the need for subwavelength arrays. 
This scaling can be directly compared with the usual requirement for Dicke superradiance in free space, i.e. atoms packed within $\sim \lambda^3$ exhibit collective behavior~\cite{dicke1954coherence, gross1982superradiance, masson2022universality, PhysRevResearch.4.023207, 38zm-mckb}; the directed emission shifts the relevant length scale from the wavelength $\lambda$ to the Rayleigh length $z_R$.

\subsection{Enhanced selective and sub-radiance in atom arrays}

Pauli blocking can be employed even in tight traps to suppress fundamental inelastic dissipation channels that degrade cooperative scattering. Indeed, previous works point out that spatial disorder degrades directed emission and subradiance in atom arrays as $\sim e^{-\eta^2}$ compared to a perfect lattice of pointlike atoms~\cite{rui2020subradiant, masson2020atomic, solomons2024universal, shahmoon2017cooperative}. If this disorder arises primarily from classical sources -- such as atomic temperature or an imperfect trapping potential -- it presents a technical challenge that can, in principle, be overcome. However, even ordered atoms cooled to the motional ground state in tight traps retain a finite zero-point size, which
fundamentally lower-bounds the inelastic scattering rate.

\begin{figure}
    \centering
    \includegraphics[width=\linewidth]{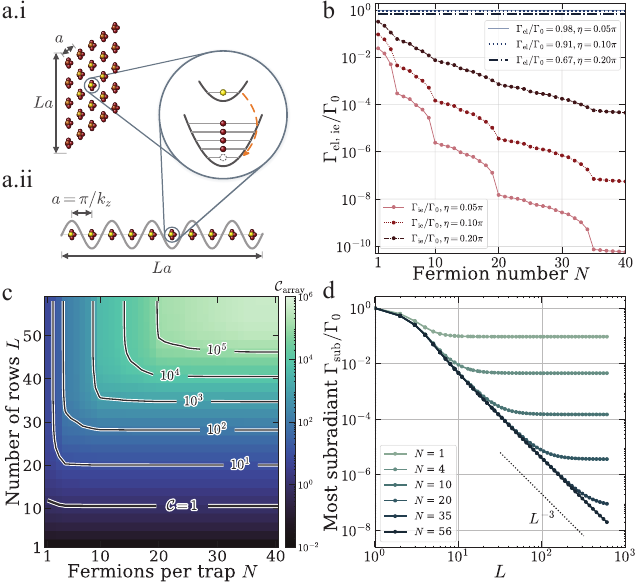}
    \caption{\textbf{Beyond Lamb-Dicke: enhanced light-matter interactions in atom arrays} (a) Sketch of 2D array of side length $La$ (i) and 1D chain of length $La$ (ii), where each site $q$ is a tight spherical trap containing $N$ fermions. (b) Elastic and inelastic scattering rates in a single spherical trap $\eta = 0.05\pi,0 .10\pi, 0.20\pi$; there is a sharp decrease in inelastic scattering at $N = 4, 10, 20, 35, ...$ whenever a new energy shell becomes fully filled. (c) Collective cooperativity for a 2D array of side length $L$ of $N$-fermion ensembles for spherical traps $\eta =\pi/10$. Without suppression of inelastic scattering, the cooperativity remains small despite the increase in the size of the grid. (d) Linewidth $\Gamma_\text{sub}$ of the most subradiant state in a 1D chain of $L$ tight traps $\eta = \pi/10$. Ideally, $\Gamma_\text{sub}$ should scale as $L^{-3}$. In practice, the linewidth saturates because it is dominated by inelastic scattering. The $L^{-3}$ scaling is recovered only when additional fermions are added to each trap.}
    \label{fig:fig5}
\end{figure}

To illustrate how Pauli blocking overcomes this limit for atom arrays, we first examine the mechanism by which inelastic scattering suppresses the cooperativity of directed emission.
Consider an $L \times L$ grid of tight spherical traps $\eta\ll 1$ in the $xy$ plane, with each site $q$ containing a single atom in its motional ground state $\ket{g,\mathbf{0}}_q$. As shown in Figure~\ref{fig:fig5}a.i, the traps are spaced by a distance comparable to the wavelength $a \lesssim \lambda$, and the total extent of the array $La \gg w_0$ is larger than the waist of target Gaussian mode $M$~[Eq.~\ref{eq:EMlobes}]. With a single photon in mode $M$, we excite a collective polarization $\hat{P}^\dagger\prod_q \ket{g,\mathbf{0}}_q$, where the operator 
\begin{equation}
    \hat{P}^\dagger = \sum_qg(s_q)\sigma_q^\dagger = \frac{ \sum_q\exp\left(-s_q^2/w_0^2\right) \sigma_q^\dagger}{\sqrt{\sum_q\exp\left(-2s_q^2/w_0^2\right)}}
\end{equation}
%
has a Gaussian envelope $g(s)$ matching the field amplitude of $\boldsymbol{\mathcal{E}}_{M}(s_q, z=0)$  at each trap center~\cite{solomons2024universal}. 

In analogy with Equation~\ref{eq:ft}, the emission pattern $\mathbf{E}(\mathbf{k})$ of the elastic scattering from the state $\hat{P}^\dagger\prod_q \ket{g,\mathbf{0}}_q$ is proportional to
\begin{equation}
    \mathbf{E}(\mathbf{k}) \propto \mathcal{F}_\text{2D}[g](\mathbf{k})\,(\mathbf{e}_x-(\mathbf{e}_x\cdot \mathbf{e}_r)\mathbf{e}_r),
\end{equation}
where $\mathcal{F}_\text{2D}[g]$ is the discrete Fourier transform of $g(s_q)$, 
and the factor $(\mathbf{e}_x-(\mathbf{e}_x\cdot \mathbf{e}_r)\mathbf{e}_r)$ ensures $\mathbf{E}(\mathbf{k})$ is transverse in the far-field. However, the state $\hat{P}^\dagger\prod_q\ket{g,\mathbf{0}}_q$ is not strictly an eigenstate of the dipole-dipole Hamiltonian of the array [Eq.~\ref{eq:Hspin}], so its emission pattern changes as it decays. We therefore calculate the total probability $P_M$ that the excitation is emitted into mode $M$ by integrating the overlap between the emitted field and the target mode (Appendix E) over time. 
Denoting the total elastic and inelastic scattering probabilities by $P_\text{el}$ and $P_\text{ie}$, with $P_\text{el}+P_\text{ie}=1$, the cooperativity is
\begin{equation}\label{eq:C_array}
    \mathcal{C}_\text{array}
    = \frac{P_M}{1-P_M}= \frac{P_M}{P_\text{el}+P_\text{ie}-P_M}.
\end{equation}

Most of the scattering into the target mode comes from elastic emission, in which the coherent interference between the emission of different trap sites produces the directed emission matched to the target mode; we denote the residual mismatch by $\varepsilon_\text{el}$. The inelastic scattering from each trap site is incoherent and its emission pattern is nearly isotropic, so its overlap with the target mode is only $\approx 0.4\%$~\cite{tanji2011interaction}; we denote this overlap by $\varepsilon_\text{ie}$. Thus $P_M = P_\text{el}(1-\varepsilon_\text{el})+P_\text{ie}\varepsilon_\text{ie}$, and the cooperativity can be written as
\begin{equation}\label{eq:C_array_epsilon}
    \mathcal{C}_\text{array} = \frac{P_M}{P_\text{el}\varepsilon_\text{el}+P_\text{ie}(1-\varepsilon_\text{ie})}.
\end{equation}

Equation~\ref{eq:C_array_epsilon} has two limits in the Lamb-Dicke regime. When the inelastic scattering probability is large, $P_\text{ie} \gg \varepsilon_\text{el}$, the cooperativity is set by $\eta^2$,
\begin{equation}\label{eq:C_array_ie}
    \mathcal{C}_\text{array} \approx \frac{P_M}{P_\text{ie}}
    \sim\frac{1}{\eta^2}.
\end{equation}
As expected, the cooperativity of the atom array is bounded by the Lamb-Dicke parameter, which in practice is typically limited to $\eta^2\gtrsim 1\%$. In the opposite limit, once Pauli blocking has closed the inelastic channels, $\Gamma_\text{ie}\rightarrow 0$ and $\mathcal{C}_\text{array}\rightarrow 1/\varepsilon_\text{el}$ is set by the mode mismatch alone.

To quantify how Pauli blocking enhances the array cooperativity $\mathcal{C}_\text{array}$, we calculate the (in)elastic scattering rates for $N$ fermions in a single spherical trap for Lamb-Dicke parameters $\eta = 0.05\pi, 0.10\pi,0.20\pi$~[Fig.~\ref{fig:fig5}b]. Because these traps are relatively tight, the elastic scattering rate is only slightly reduced from the intrinsic linewidth, $\Gamma_\text{el} \approx \Gamma_0$ (blue lines). For each value of $\eta$, the inelastic scattering rate drops sharply each time an energy shell of the trap is filled (red curves). Specifically, filling the $n^\text{th}$ shell reduces the inelastic scattering by $e^{-\eta^2}\eta^{2n}/n!$, as successive orders in the expansion $\Gamma_\text{ie} = (1 - e^{-\eta^2})\Gamma_0 =  e^{-\eta^2}( \eta^2 + \eta^4/2! + \ldots)\Gamma_0$ are blocked.

Assuming that all traps in the array are filled identically up to the $n_0^\text{th}$ shell, the inelastic scattering rate is
\begin{equation}\label{eq:Gamma_ie_shells}
    \Gamma_\text{ie} = \Big(1 - e^{-\eta^2}\sum_{n=0}^{n_0}\frac{\eta^{2n}}{n!}\Big)\Gamma_0,
\end{equation}
and $\mathcal{C}_\text{array}$ follows from Eq.~\ref{eq:C_array}. Figure~\ref{fig:fig5}c shows the result for a square grid of $L\times L$ traps with $\eta=\pi/10$, spacing $a=0.6\lambda$, and a target mode of waist $w_0=6\lambda$ in the atom plane. As the array grows, the cooperativity should ideally scale as $\mathcal{C}_\text{array}\sim L^4/(\log L)^2$ for an optimally chosen waist, as the mode matching improves~\cite{manzoni2018optimization}. Instead, it is dominated by inelastic scattering and remains below $10$ for all array sizes $L$ in the $N=1$ column. Only when fermions are added to each trap to block the inelastic scattering does the cooperativity exhibit the expected scaling with $L$.

Similarly, Pauli blocking can enhance subradiance within a one-dimensional chain of $L$ tight spherical traps where the disorder is dominated by the Lamb-Dicke parameter~[Fig.~\ref{fig:fig5}a.ii]. For example, Ref.~\cite{asenjo2017exponential} showed that the linewidth of the most subradiant mode for such a chain should scale as $\Gamma_\text{sub} \sim \Gamma_0/L^3$. At fixed Lamb-Dicke parameter, however, this is not the case; the minimum linewidth saturates at $\Gamma_\text{ie}$~[Eq.~\ref{eq:Gamma_ie_shells}] [Fig.~\ref{fig:fig5}d, $N=1$ curve]. Thus the effect of the zero-point size is to directly limit the subradiance, setting a minimum scattering rate which is independent of length $L$, spacing $a$, or wavevector $k_z$. However, adding fermions to each trap decreases the inelastic scattering as shown in Figure~\ref{fig:fig5}b, such that the $L^{-3}$ scaling is recovered as shown in Figure~\ref{fig:fig5}d.

More broadly, isolating elastic scattering in such a system slightly modifies the dipole-dipole interactions. Compared to the idealized point-atom case, our model accounts for the finite Gaussian spread of the atomic wavefunctions and the presence of multiple fermions per trap. The finite wavefunction modifies the point-like dipole-dipole interactions, scaling $J$ and $\gamma$ by $e^{-\eta^2}$. Similarly, the Fermi sea modifies the single-atom jump rate to 
$\Gamma_0 e^{-\eta^2}$. Thus, the effective Hamiltonian and the jump terms undergo a uniform rescaling by $e^{-\eta^2}$, which maps directly onto a globally scaled band structure.

\subsection{Arbitrarily directed emission}\label{sec:uni}

Extending the framework established in the previous sections, we now introduce a method to tailor elastic emission patterns arbitrarily through the application of external driving fields to multi-level atoms. For instance, our scheme can produce unidirectional spontaneous emission, which is non-trivial as it necessitates breaking time-reversal symmetry. In past work, uni-directional emission has been achieved via spin-orbit interactions or the Kerker effect~\cite{liu2018generalized, poshakinskiy2019optomechanical, rodriguez2013near}, concepts which have been applied to cold atom systems theoretically 
~\cite{alaee2020kerker, PhysRevLett.125.143604, lodahl2017chiral, jones2020collectively, pichler2015quantum, masson2020atomic} and experimentally~\cite{PhysRevX.5.041036, mitsch2014quantum, pucher2022atomic, PhysRevX.14.011020}. Our method is distinct from these approaches in that it harnesses the elastic scattering amplitude of Equation~\ref{eq:ft}, but breaks the symmetry $A_\mathbf{00}(\mathbf{k}) = A_\mathbf{00}(-\mathbf{k})$ imposed by the real atomic density $|\psi_\mathbf{0}(\mathbf{r})|^2$.

\begin{figure}
    \centering
    \includegraphics[width=\linewidth]{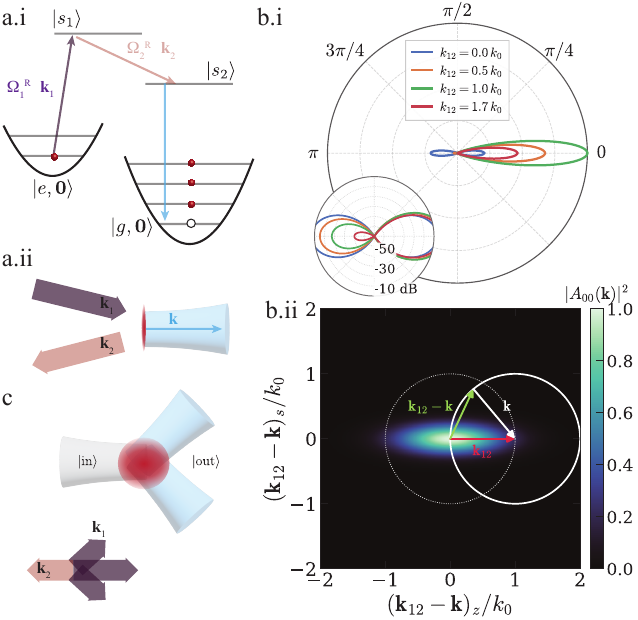}
    \caption{\textbf{Arbitrary emission via an external drive:} (a) Four internal states $e, s_1, s_2, g$, coupled by two external driving fields $\Omega_1^\text{R}, \mathbf{k}_1$ and $\Omega_2^\text{R}, \mathbf{k}_2$ (i), which induce a wavepacket antenna (red) to emit to the right (blue) when arranged as shown with $\mathbf{k}_{12}$ along $+z$. (ii). (b) Effect of $\mathbf{k}_{12}$ on spontaneously emitted photons $\ket{s_2} \rightarrow \ket{g,\mathbf{0}}$. Radar plot (i) shows the unidirectional emission increasing to the right as $\mathbf{k}_{12}$ increases, with inset showing log scale. The effect of $\mathbf{k}_{12}$ is to shift the allowed momentum to the right in $k$-space (ii). When $\mathbf{k}_{12}$ is zero, photons are emitted on the dotted white circle; the transition probabilities $|A_\mathbf{00}(\mathbf{k})|^2$ are symmetric. The drive $\mathbf{k}_{12}$ shifts the allowed emission wavevectors to the right, so the transition probability $|A_\mathbf{00}(\mathbf{k})|^2$ projected onto the solid white circle is asymmetric. (c) A spherical Gaussian wavepacket atom under a shaped driving field $\mathbf{k}_1, \mathbf{k}_2$ as shown deterministically converts single photons from a Gaussian input mode into a double-lobed output.}
    \label{fig:fig6}
\end{figure}

In particular, we engineer an asymmetric elastic transition amplitude $A_\mathbf{00}(\mathbf{k})$ in Fourier space, using an external drive to break the time-reversal symmetry. Starting from the initial state $\ket{\psi_N'}$ for a single trap, %
%
we now assume 
that the excited fermion decays via the pathway $\ket{e}\leftrightarrow  \ket{s_1} \leftrightarrow \ket{s_2} \rightarrow \ket{g}$, where $\ket{s_1}$ and $\ket{s_2}$ are intermediate states and $\ket{e}$ is now metastable.
As shown in Figure~\ref{fig:fig6}a, the first two transitions in the decay process are driven via coherent control fields: the first resonantly couples $\ket{e}\leftrightarrow\ket{s_1}$ with Rabi frequency $\Omega^\text{R}_1$ and wavevector $\mathbf{k}_1$ (purple); the second resonantly couples $\ket{s_1}\leftrightarrow\ket{s_2}$ with Rabi frequency $\Omega^\text{R}_2$ and wavevector $\mathbf{k}_2$ (pink). The final leg is a spontaneous decay event $\Gamma_{s_2g}$, which exhibits the desired unidirectional emission (blue). With $\Omega^\text{R}_2 \gg \Omega^\text{R}_1, \Gamma_{s_1e}$ and $\Gamma_{s_2g}\gtrsim |\mathbf{\Omega}_{s_2}|, |\mathbf{\Omega}_e|, \hbar k^2_{12}/2m$, the states $\ket{s_1, \mathbf{n}}$ and $\ket{s_2, \mathbf{n}}$ can be adiabatically eliminated (Appendix E).

To understand the unidirectional emission pattern of the final leg, we observe that an atom initially in $\ket{e,\mathbf{0}}$ can only decay to $\ket{g,\mathbf{0}}$ by absorbing a drive photon with momentum $\mathbf{k}_1$, then emitting a photon with momentum $\mathbf{k}_2$ via stimulated emission. The final spontaneous emission momentum $\mathbf{k}$ into a random direction yields a total recoil momentum  $\mathbf{k}_\text{total} = \mathbf{k}_1 - \mathbf{k}_2 - \mathbf{k} = \mathbf{k}_{12} - \mathbf{k}$. 
Thus, the transition amplitudes $A_\mathbf{nm}(\mathbf{k})$~[Eq.~\ref{eq:Anm}] are modified,
and the elastic scattering amplitude becomes
\begin{equation}\label{eq:A00shift}
A_\mathbf{00}(\mathbf{k}_\text{total}) = \bra{ \mathbf{0} } e^{i(\mathbf{k}_{12} - \mathbf{k})\cdot \hat{\mathbf{r}}}\ket{\mathbf{0}} = \mathcal{F}[|\psi_\mathbf{0}|^2](\mathbf{k}-\mathbf{k}_{12}),
\end{equation}
which corresponds to a translation in Fourier space.

Figure~\ref{fig:fig6}b illustrates a concrete example of using momentum translation to produce asymmetric Gaussian emission. The elastic scattering intensity for light spontaneously emitted on the $\ket{s_2} \rightarrow \ket{g}$ transition is plotted in Figure~\ref{fig:fig6}b.i for a disk emitter $\eta_s = \sqrt{2}\pi, \eta_z = 2\pi/5$. 
The radius is $(1-|\mathbf{e_d}\cdot\mathbf{e_k}|^2)\,|A_\mathbf{00}(\mathbf{k}-\mathbf{k}_{12})|^2$, which reaches unity at exact momentum matching $\mathbf{k}=\mathbf{k}_{12}$. For $\mathbf{k}_{12}=0$, the intensity is symmetric with a magnitude reduced from the maximum as $e^{-\eta_z^2}$~[Eq.~\ref{eq:Gamma_el/Gamma_0}] (blue curve).  
However, choosing the orientation of $\mathbf{k}_{12}$ along the $+z$-axis results in preferential emission to the right as $k_{12}$ increases from zero to $0.5k_0, 1.0k_0, 1.7k_0$ (orange, green, red curves, respectively).

Figure~\ref{fig:fig6}b.ii illustrates this process explicitly in $k$-space. When $\mathbf{k}_{12}=0$, conservation of momentum restricts photon recoil to wavevectors lying on the dotted white circle, with an intensity given by the colorscale $|A_\mathbf{00}(\mathbf{k})|^2$. For the disk emitter, $A_\mathbf{00}(k_s, k_z)$ is maximized at $k_s = 0$ and is symmetric for $\pm k_z$. Applying the momentum $\mathbf{k}_{12}$ translates the allowed wavevectors for $\ket{s_2}\rightarrow \ket{g}$ emission, shown in Figure~\ref{fig:fig6}b.ii as the solid white circle shifted to the right. While the recoil amplitude $A_{\mathbf{00}}$ remains unchanged, its overlap with the translated circle becomes asymmetric. In particular, the amplitude is enhanced for recoil to the left (photon emission to the right), and suppressed in the opposite direction. The enhancement is maximized for $k_{12} = k_0$.

To move beyond unidirectional emission, we relax the assumption that the driving fields are plane waves and allow them to vary spatially, with dimensionless profiles $\mathcal{E}_1(\mathbf{r})$ and $\mathcal{E}_2(\mathbf{r})$ normalized to unit peak amplitude. Adiabatically eliminating the intermediate states $\ket{s_1}$ and $\ket{s_2}$ then replaces the recoil phase $e^{i\mathbf{k}_{12}\cdot\hat{\mathbf{r}}}$ by the ratio of the two profiles~(Appendix~E), and the elastic transition amplitude becomes
\begin{equation}\label{eq:A00shaped}
    A_\mathbf{00}(\mathbf{k}) = \Big\langle \mathbf{0}\Big|\,
    \frac{\mathcal{E}_1(\hat{\mathbf{r}})}{\mathcal{E}_2(\hat{\mathbf{r}})}\,
    e^{-i\mathbf{k}\cdot\hat{\mathbf{r}}}\,\Big|\mathbf{0}\Big\rangle.
\end{equation}
The radiation pattern can therefore be tailored by shaping the drives; for example, if $\mathcal{E}_1(\mathbf{r})$ consists of three plane waves traveling in different directions, the radiation pattern has three lobes. As shown in Figure~\ref{fig:fig6}c, by correctly choosing the directions and relative powers of the three plane waves, one can engineer destructive interference with the input beam such that the output light is emitted in only two lobes (Appendix~E). 
Thus, the protocol enables nonlinear modification of the spatial mode of a single photon.

The same protocol applied to multiple emitters yields a flexible method for modifying the dipole-dipole interactions between them. In the single-excitation manifold, the drives dress the chain Hamiltonian of Equation~\ref{eq:Hspin} on both sides,
\begin{equation}\label{eq:dressedH}
    H_\text{1D} \rightarrow \Lambda^\dagger H_\text{1D} \Lambda ,
    \qquad \Lambda = \text{diag}(\hat{\mathcal{R}}_1, \hat{\mathcal{R}}_2, \ldots) ,
\end{equation}
with $\hat{\mathcal{R}}_q \approx -(\Omega^\text{R}_1/\Omega^\text{R}_2)\,\mathcal{E}_1(\hat{\mathbf{r}}_q)/\mathcal{E}_2(\hat{\mathbf{r}}_q)$ the two-photon coupling at site $q$~(Appendix~E). For plane-wave drives, $\hat{\mathcal{R}}_q = -(\Omega^\text{R}_1/\Omega^\text{R}_2)\, e^{i\mathbf{k}_{12}\cdot\hat{\mathbf{r}}_q}$, so $\Lambda$ is unitary up to the small factor $\Omega^\text{R}_1/\Omega^\text{R}_2$: every rate and coupling is reduced by $|\Omega^\text{R}_1/\Omega^\text{R}_2|^2$, while the phase offsets the recoil momentum by $\mathbf{k}_{12}$ and translates the radiation pattern of every mode of the chain in momentum space. For e.g. the chain of Figure~\ref{fig:fig4}c, the entire band structure is translated, $k_z \rightarrow k_z - k_{12,z}$, with no change to the emitter spacing $a$. 
Shaping $\mathcal{E}_1$ and $\mathcal{E}_2$ modifies the coupling site by site. Because the dressed emitter radiates asymmetrically, the coupling from site $q$ to $q'$ differs in magnitude from its reverse and can be used to realize cascaded quantum systems~\cite{PhysRevLett.70.2269, PhysRevLett.70.2273}.

\section{Conclusion and outlook}
We have introduced a number of directions in quantum optics using highly controlled fermionic ensembles. Beginning with a method to produce directed emission from a single atomic wavepacket at the bottom of a Fermi sea, we studied dipolar interactions, suppression of inelastic scattering, and arbitrary emission for multi-level systems. Our work opens the door to intriguing opportunities involving the interplay of quantum statistics and spatially ordered atomic dipoles, such as moving beyond the single-excitation subspace, controlling spin-exchange processes within and between traps, and realizing these protocols experimentally.


\begin{acknowledgments}
We gratefully recognize stimulating and helpful conversations with James Thompson, David Weld, and Dan Stamper-Kurn. We thank Francisco Machado and Avikar Periwal for insightful comments on the manuscript. 
\end{acknowledgments}

\bibliography{references}
\end{document}